\documentclass[a4paper,11pt]{article}
\usepackage{jinstpub}
\usepackage{microtype}
\usepackage{booktabs}
\usepackage{multirow}
\usepackage{subcaption}
\usepackage{float}
\usepackage{arydshln}

\title{Reflectance and fluorescence characteristics of PTFE coated with TPB at visible, UV, and VUV as a function of thickness}

\author[10,a]{J.~Haefner,\note[a]{Corresponding author.}}
\author[10,a]{A.~Fahs,}
\author[18]{J.~Ho,}
\author[18]{C.~Stanford,}
\author[18]{R.~Guenette,}
\author[2]{C.~Adams,}
\author[18]{H.~Almaz\'an,}
\author[26]{V.~\'Alvarez,}
\author[23]{B.~Aparicio,}
\author[22]{A.I.~Aranburu,}
% ...
\affiliation[2]{
Argonne National Laboratory, Argonne, IL 60439, USA}
% ...
\affiliation[10]{
Department of Physics, Harvard University, Cambridge, MA 02138, USA}
% ...
\affiliation[18]{
Department of Physics and Astronomy, Manchester University, Manchester. M13 9PL, United Kingdom}
% ...
\affiliation[22]{
Department of Applied Chemistry, Universidad del Pais Vasco (UPV/EHU), Manuel de Lardizabal 3, San Sebasti\'an / Donostia, E-20018, Spain}
\affiliation[23]{
Department of Organic Chemistry I, University of the Basque Country (UPV/EHU), Centro de Innovaci\'on en Qu\'imica Avanzada (ORFEO-CINQA), San Sebasti\'an / Donostia, E-20018, Spain}
% ...
\affiliation[26]{
Instituto de Instrumentaci\'on para Imagen Molecular (I3M), Centro Mixto CSIC - Universitat Polit\`ecnica de Val\`encia, Camino de Vera s/n, Valencia, E-46022, Spain}
% ...

\abstract{Polytetrafluoroethylene (PTFE) is an excellent diffuse reflector widely used in light collection systems for particle physics experiments. In noble element systems, it is often coated with tetraphenyl butadiene (TPB) to allow detection of vacuum ultraviolet scintillation light. In this work this dependence is investigated for PTFE coated with TPB in air for light of wavelengths of 200~nm, 260~nm, and 450~nm. The results show that TPB-coated PTFE has a reflectance of approximately 92\% for thicknesses ranging from 5~mm to 10~mm at 450~nm, with negligible variation as a function of thickness within this range. A cross-check of these results using an argon chamber supports the conclusion that the change in thickness from 5~mm to 10~mm does not affect significantly the light response at 128~nm. Our results indicate that pieces of TPB-coated PTFE thinner than the typical 10~mm can be used in particle physics detectors without compromising the light signal.}


\arxivnumber{1234.56789} 

\begin{document}
\maketitle
\flushbottom

\section{Introduction} \label{sec:Introduction}

Polytetrafluoroethylene (PTFE), often referred to by the brand name \textsc{Teflon}, is a plastic with excellent diffuse reflectance and low cost. These properties make it commonly used in particle physics experiments (e.g., \cite{Alvarez:2012, Auger:2012, Akerib:2015}) for light collection. The NEXT experiment uses PTFE as a diffuse reflector around the drift volume of its high pressure gaseous xenon time projection chamber (TPC) to increase light collection \cite{Alvarez:2012}, in an attempt to search for neutrino-less double beta decay. The measurements described here have informed the design decisions for the upcoming phases of NEXT, including NEXT-100~\cite{Martin-Albo:2015} (where the thickness of PTFE has been halved relative to previous detectors) and a future tonne-scale experiment~\cite{Adams:2020}.  

While PTFE is excellent for light collection, its radiopurity can be a concern for low-background experiments~\cite{Novella:2019}, especially at large scale, where a significant quantity of material is needed. Moreover, PTFE can absorb gaseous xenon~\cite{Rogers:2018} leading to a loss of active detector mass, and outgas contaminants in the high purity detector medium. Minimizing the quantity of PTFE in such experiments is therefore an attractive solution, provided that the light collection remains high. 

A significant number of particle physics experiments that use PTFE, especially for rare event searches, utilize liquid or gaseous noble element detectors. These detectors typically collect the scintillation light, produced when a particle interacts in the detection medium. Noble elements scintillate primarily in the vacuum ultraviolet (VUV) range. At these wavelengths, common photosensors ---such as silicon photomultipliers (SiPMs) and photomultiplier tubes (PMTs)--- generally have very low sensitivity and wavelength shifters are often used to shift the VUV light to the visible range. Similarly to many comparable experiments, the NEXT experiment uses tetraphenyl butadiene (TPB) to shift the 175~nm light produced in xenon to the visible $\sim$420~nm (blue). The light response properties of PTFE in these regimes, both at VUV and in blue, are thus of great interest to many particle physics experiments.

A variety of studies have been carried out in the past to investigate reflectance of PTFE, such as the reflectance of PTFE at 175~nm at a single 5~mm thickness~\cite{Silva:2009}, reflectance in liquid xenon for thicknesses from 1~mm to 9.5~mm~\cite{Haefner:2016}, reflectance of PTFE thicknesses from 5~mm to 10~mm for visible and UV light~\cite{Ghosh:2020}, and reflectance as a function of angle~\cite{Kravitz:2020}.

The light response of PTFE coated with a wavelength shifter such as TPB is less explored, although there have been several studies on the fluorescence of TPB itself ~\cite{Burton:1973, Benson:2018}. In the case of TPB-coated PTFE, fluorescence must also be accounted for and pure reflectance no longer gives a reasonable description of the overall light response behavior in the UV. In this work, the variation of light response of TPB-coated PTFE is investigated for a variety of wavelengths ranging from VUV to blue, as a function of PTFE thickness using four distinct methods.

The first two methods involve measuring TPB-coated PTFE disks in commercial spectrophotometers. These are devices which are commonly used for materials characterization in which light of one or several wavelengths is incident upon a given sample to measure a variety of optical properties, such as fluorescence or reflectance. In Section~\ref{ssec:method1}, reflectance spectra of TPB-coated and uncoated PTFE disks of different thicknesses are measured using a universal measurement spectrophotometer and in Section~\ref{ssec:method2}, fluorescence spectra of the same disks are measured using a fluorescence spectrometer.

The remaining two methods involve enclosed boxes made of PTFE of different thicknesses, where the reflectance is studied by measuring the light intensity as a function of the distance of our light source and detector plane from the end of the box. These measurements are made in two ways where Section~\ref{ssec:method3} describes measurements in air with an LED light sources at 260~nm and 450~nm, and Section~\ref{ssec:method4} describes measurements in an argon gas chamber measuring scintillation light at 128~nm from alpha decay events from a $^{210}$Po source.

\section{Sample Preparation} \label{sec:Measurements}
In order to investigate the reflectance of PTFE, several $12~\mathrm{in}\times24~\mathrm{in}$ sheets of different thicknesses were purchased. All sheets for the primary measurements were sourced from \textsc{Altec Plastics}.\footnote{https://altecplastics.com/} The different thicknesses are $\frac{3}{8}$~in = 9.5~mm, $\frac{5}{16}$~in = 7.9~mm, $\frac{1}{4}$~in = 6.4~mm, and $\frac{3}{16}$~in = 4.8~mm, referred to as 10~mm, 8~mm, 6~mm, and 5~mm respectively. Additional sheets were procured from \textsc{Industrial Plastics}\footnote{https://www.industrialplasticsinc.com/} (thicknesses of 9.8~mm, 7.7~mm, 7.2~mm and 5~mm ) and \textsc{ePlastics}\footnote{https://www.eplastics.com/} (thicknesses of 9.5~mm, 6.5~mm, 5~mm, and 3.4~mm) in order to assess variation between different sheets and manufacturers. This information is summarised in Table \ref{tab:thickness}. All the parts used for the studies, such as the disks in Sections \ref{ssec:method1} and \ref{ssec:method2} as well as the boxes in \ref{ssec:method3} and \ref{ssec:method4} were cut from these same sheets.

\begin{table}[!htb]
    \caption{Summary of the various PTFE sheet thicknesses (in mm) studied by manufacturer.}
    \centering
    \begin{tabular}{|c|c|c|}
     \hline
    \textsc{Altec Plastics} & \textsc{Industrial Plastics} & \textsc{ePlastics} \\
    \hline
    9.5 & 9.8 & 9.5 \\
    7.9 & 7.7 & 6.5\\
    6.4 & 7.2 & 5 \\
    4.8 & 5 & 3.4\\
     \hline
    \end{tabular}
    \label{tab:thickness}
\end{table}

All the pieces cut from the PTFE sheets were treated and cleaned in one day in the same way based on methods commonly used in particle physics experiments~\cite{Haefner:2016}. One side of each piece was sanded by hand with 300-, then 1000-, and finally 2000-grit sandpaper to make the surface uniform and matte. Each piece was then cleaned in an ultrasonic bath in a solution of deionized (DI) water and Alconox for 15 minutes, followed with an ultrasonic bath run in pure DI water for 15 minutes and the pieces were air-dried overnight. Following an initial run of measurements described in the next sections of all the bare (uncoated) PTFE pieces, the pieces were evaporation-coated with a layer of TPB. The thickness of this layer was measured by a crystal monitor, but since the monitor can only measure at one point, it does not necessarily represent a uniform layer. Across the different evaporations, the thickness values read by the crystal monitor ranged from $25.3 ~\mu$m to  $26.3 ~\mu$m.  In order to check for the level of non-uniformities in the evaporator, a coating test was performed with glass slides at a variety of positions in the same plane in the evaporator, where the weights of the slides were measured before and after evaporation to the precision level of $0.1$~milligrams. In the area of the evaporator where the PTFE pieces are coated for the studies, a maximum difference of $23.2~\mu$m to $40.5~\mu$m ($42\%$ difference) was observed in the coating thickness from the center of the area to the edge. To convert from the mass measurement to thickness, a uniform layer of TPB was assumed, as the surface is small enough (~1~in). This could correspond to a variance in reflectance of up to 15\%-20\%, depending on the wavelength \cite{Benson:2018}.
In order to reduce the level of non-uniformities observed with the test slides, the PTFE pieces were positioned in a way to be perpendicular to the TPB crucible, as opposed to in the plane parallel to the top of the evaporator as was done with the test slides. All the pieces of PTFE were placed equidistant from the TPB crucible, meaning that the coating on each piece should be identical. On the disks, due to their small diameter and small area investigated, non-uniformities are not expected to have a significant impact on the reported reflectances. For the box method, concerns of non-uniformity are addressed by the systematic errors.  
All PTFE pieces were coated in the same week to minimize any potential relative degradation between the pieces. As UV light is known to degrade TPB coatings ~\cite{Jones:2013}, and our previous study also suggested a possible degradation of light response for exposed uncoated PTFE ~\cite{Ghosh:2020}, all samples were stored in a cabinet wrapped in aluminium foil, protected from light. 

\begin{figure}[!htb]
    \centering
    \includegraphics[width=0.4\textwidth]{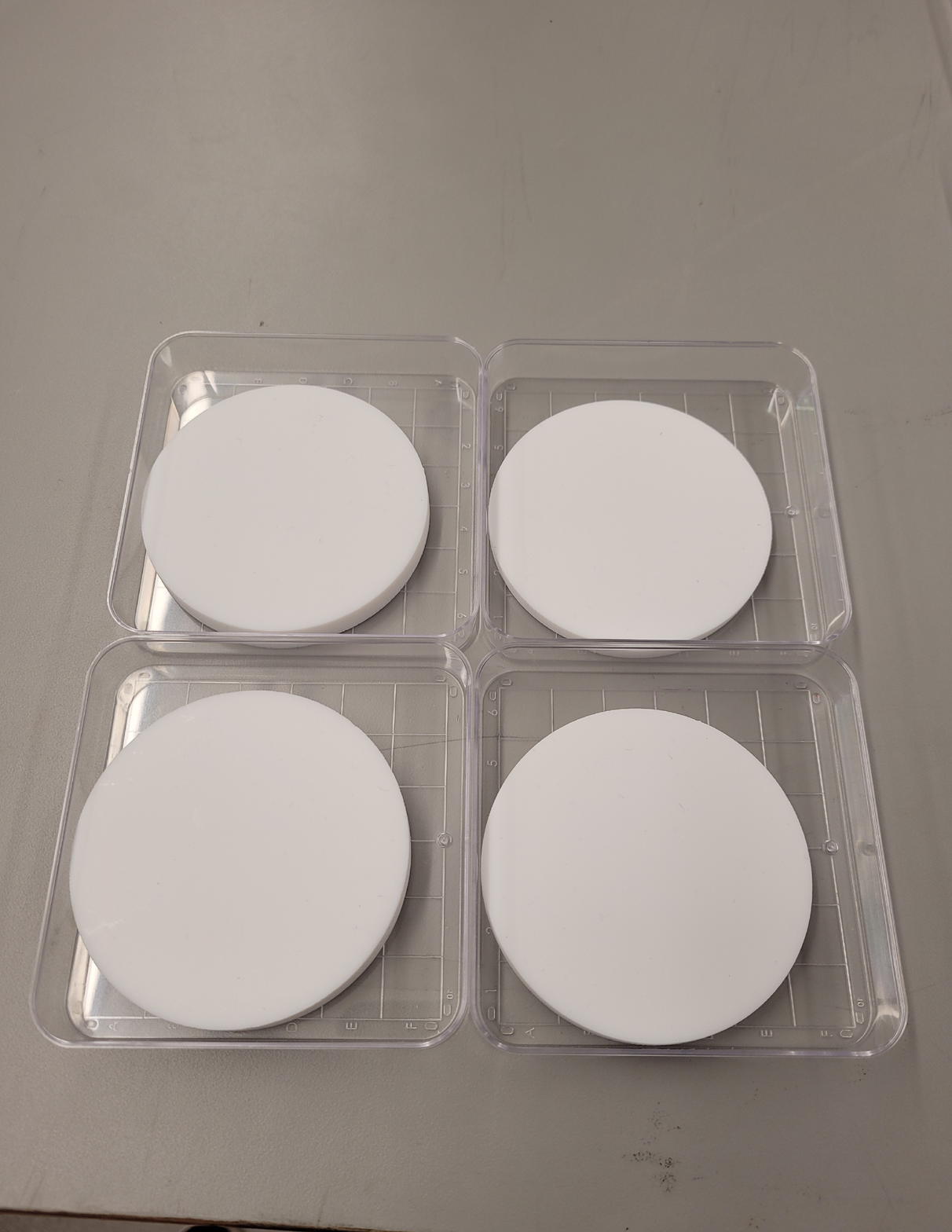}
    \includegraphics[width=0.4\textwidth]{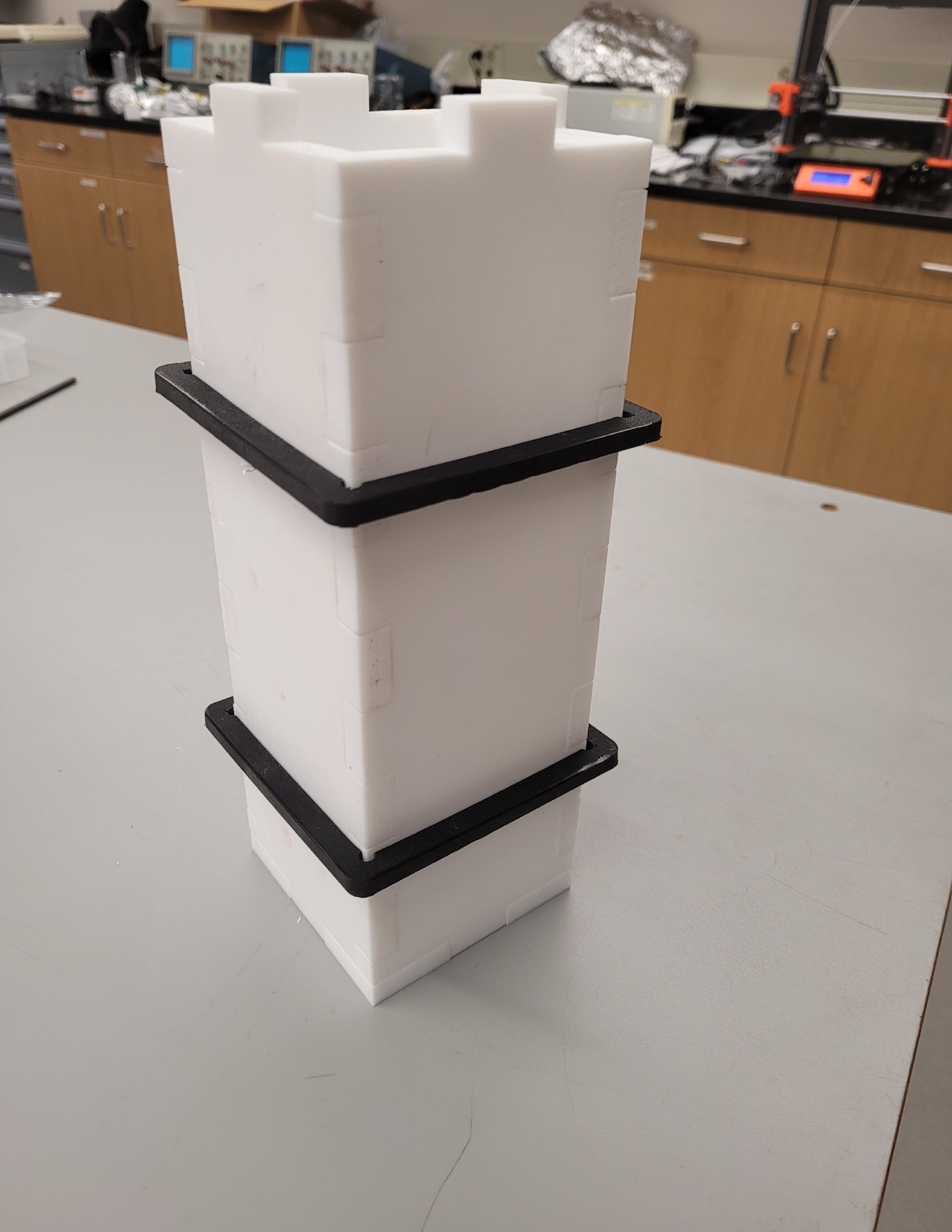}
    \caption{Left: 7.6~cm diameter PTFE disks used for measurements in the spectrophotometers. Right: An example of a PTFE box used for reflectance measurements. The boxes have inner dimensions of $7~\mathrm{cm}\times7~\mathrm{cm}\times25~\mathrm{cm}$.}
    \label{fig:PTFE_pieces}
\end{figure}

Figure~\ref{fig:PTFE_pieces} shows the different PTFE pieces used in the measurements presented here. The disks, of 7.6~cm diameter, are used for measurements in the spectrophotometers (Sections~\ref{ssec:method1} and \ref{ssec:method2}) and the box with inner dimensions $7~\mathrm{cm}\times7~\mathrm{cm}\times25~\mathrm{cm}$ are used for reflectance measurements (Sections~\ref{ssec:method3} and \ref{ssec:method4}).

\section{Universal measurement spectrophotometer} \label{ssec:method1}

\begin{figure}[tb]
\centering
\includegraphics[width=0.33\textwidth]{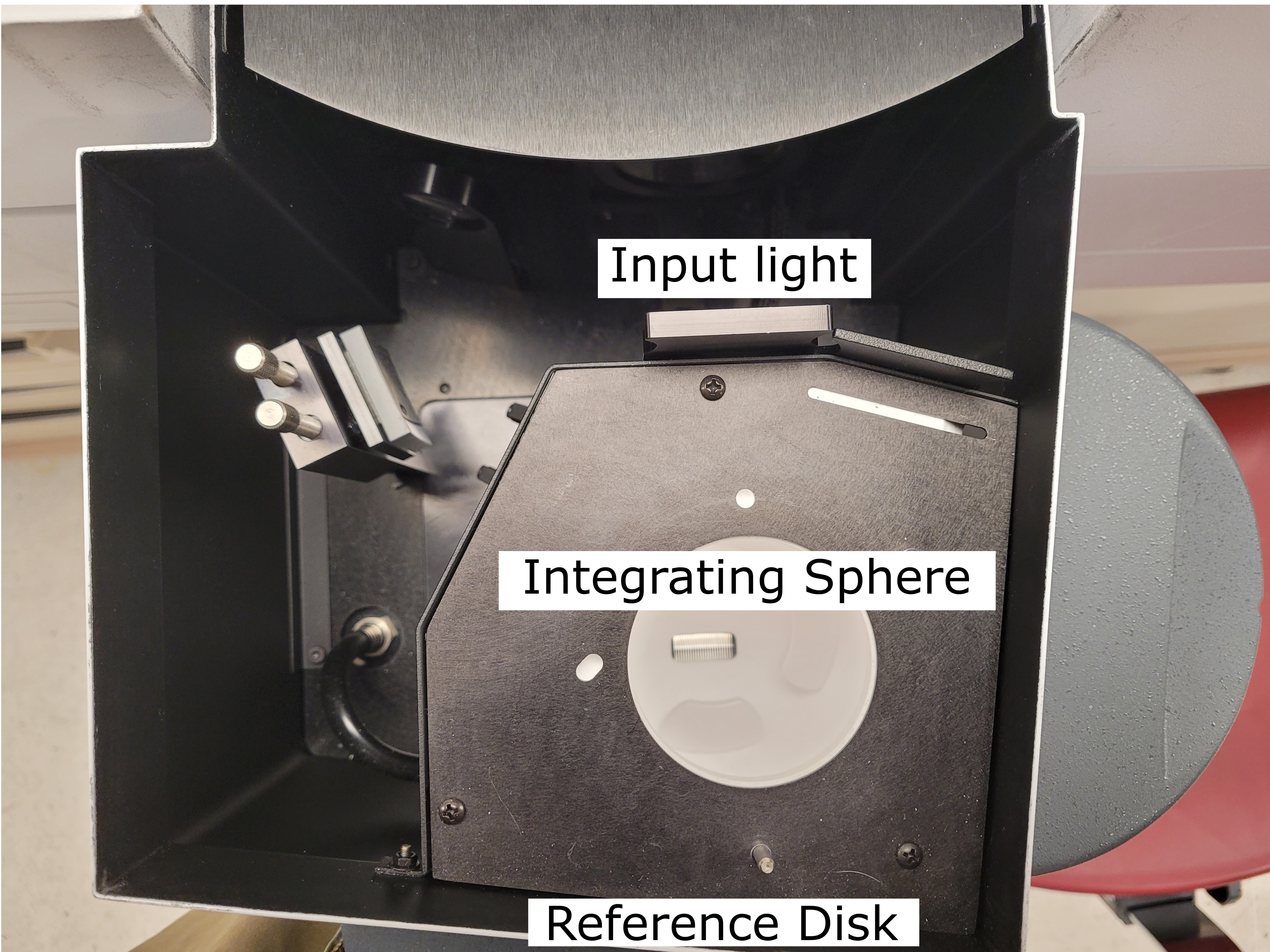}
\caption{The \textsc{Cary 7000 Spectrophotometer} with the internal components of the integrating sphere exposed.}

\label{fig:photometer_with_shroud}
\end{figure}

The first method for determining reflectance makes use of \textsc{Cary 7000 Universal Measurement Spectrophotometer} (UMS) coupled to an \textsc{Agilent} diffuse reflectance integrating sphere,\footnote{https://www.agilent.com/cs/library/flyers/public/5991-1717EN\_PromoFlyer\_UV\_DRA.pdf} shown in Figure~\ref{fig:photometer_with_shroud}. 
The UMS allows to measure the reflectance of a given sample relative to some reference, as a function of wavelength. The measurements presented here are taken relative to a Labsphere Certified Reflectance Standard (a Spectralon\textsuperscript{\textregistered} puck) with $99\%$ uniform reflectance in the 400 to 500 ~nm range, and $>95\%$ reflectance in the 250 to 400~nm range.

\begin{figure}[!htb]
\centering
\includegraphics[width=0.65\textwidth]{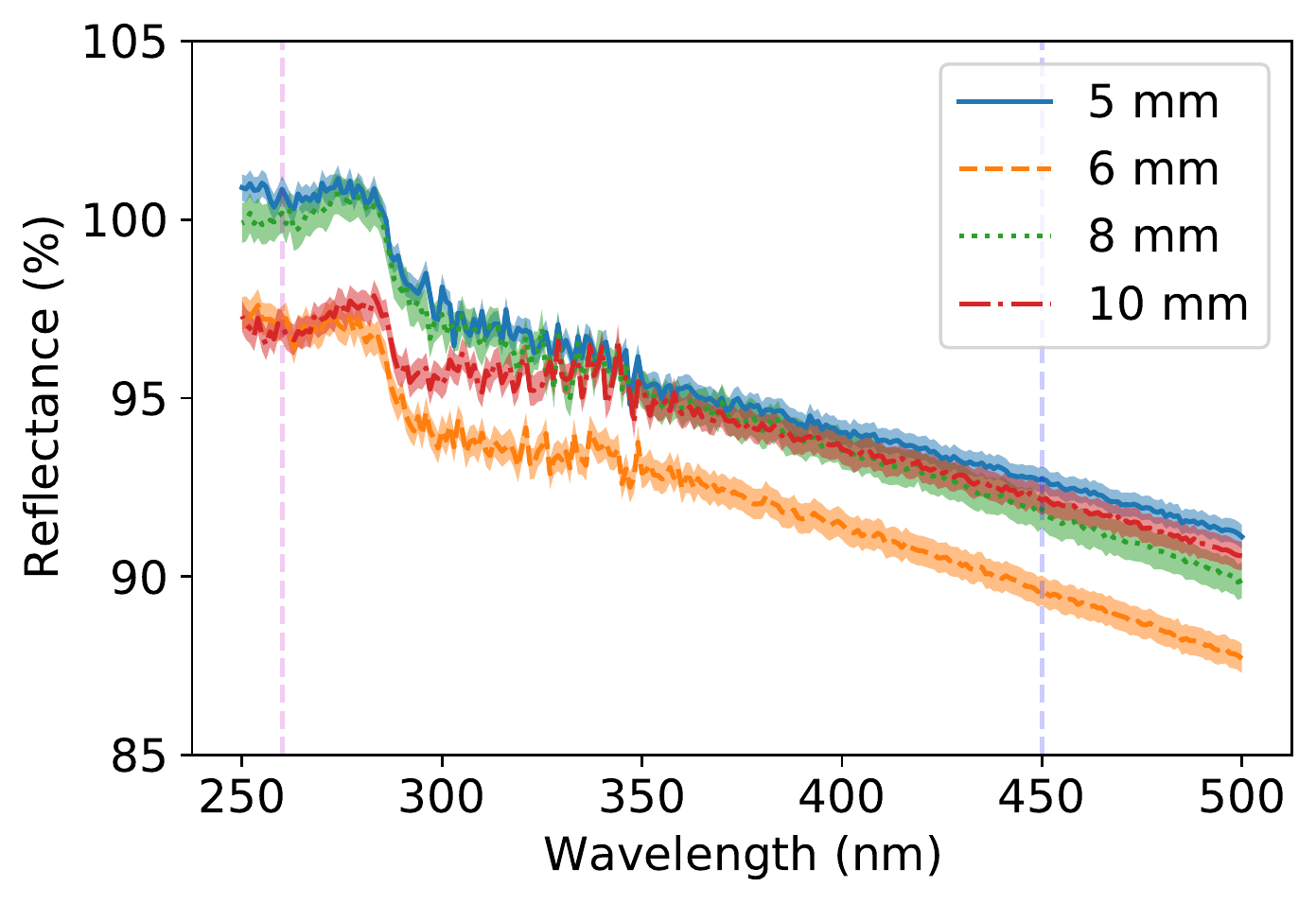}
\caption{Example of reflectance measurements from the Universal Measurement Spectrophotometer for different uncoated \textsc{Altec} PTFE thicknesses (colored lines) relative to a certified reflectance Spectralon reference, which corresponds to 100\% reflectance on the vertical axis. The purple and blue dashed lines represent the wavelengths that are used for comparisons with other methods. }
\label{fig:spm}
\end{figure}



The UMS is used to measure the reflectance of the both the uncoated and TPB-coated PTFE disks of varying thicknesses, from all the manufacturers.  Figure~\ref{fig:spm} shows an example of the measurements of the reflectance of the different \textsc{Altec} PTFE disk thicknesses relative to the Spectralon standard. Since this is a relative measurement, reflectance values over 100\% indicate the sample is more reflective than the Spectralon standard. The error bars are estimated from taking twenty measurement runs of each disk. As seen in Figure \ref{fig:spm}, the differences between thicknesses only rise to the few percent level at most, but the variations are dependant on wavelength. Significant variability was observed in reflectance across disks of the same thickness, especially when the disks were cut from different sheets.

\subsection{Results from the universal measurement spectrophotometer}
\label{sec:spm_results_section}

\begin{figure}[!htb]
\centering
\includegraphics[width=0.70\textwidth]{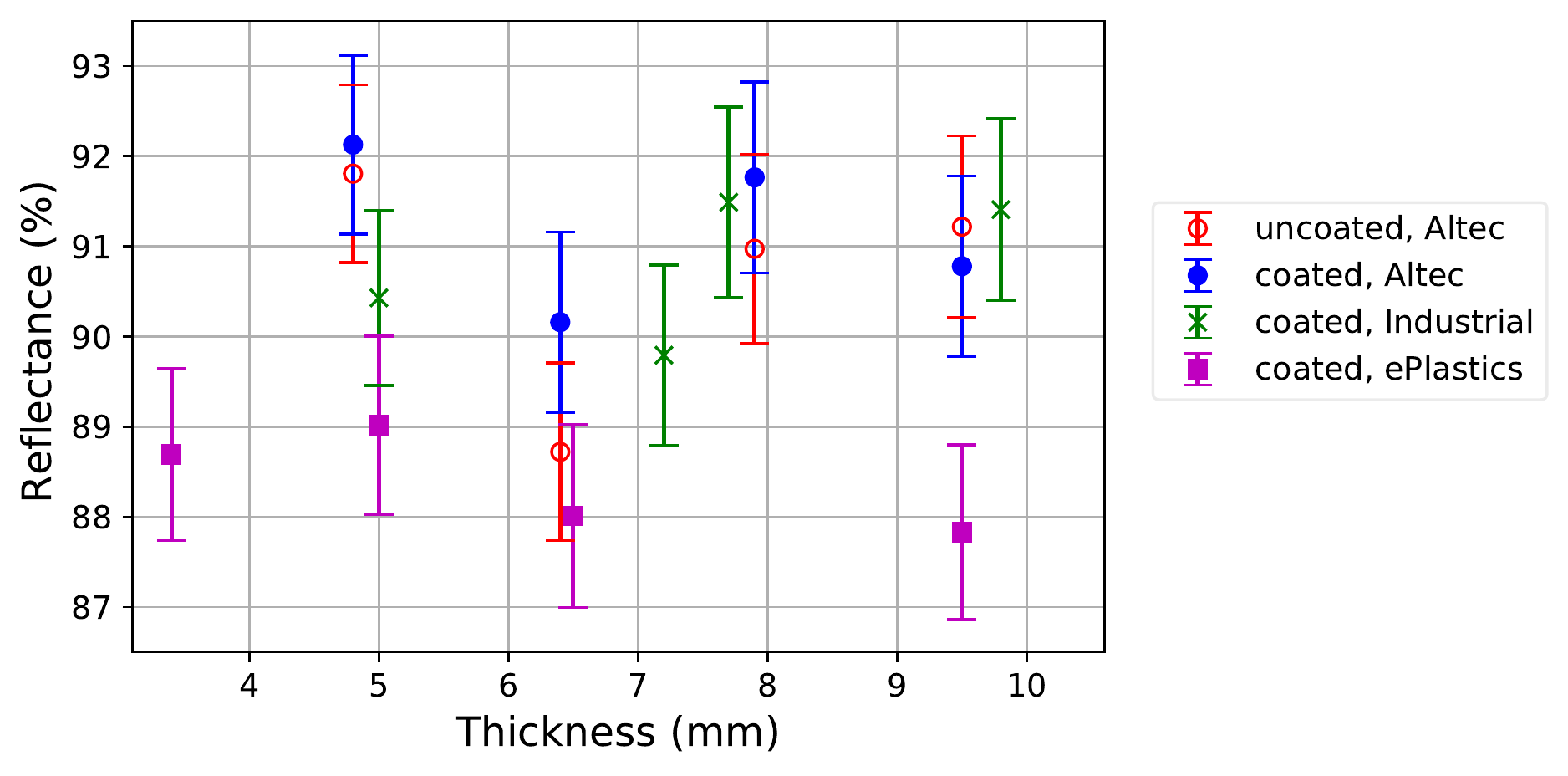}
\caption{Reflectance measurements from the Universal Measurement Spectrophotometer for different thicknesses of TPB-coated PTFE (thicknesses are shown on the horizontal axis) at 450~nm. Measurements were taken relative to the certified $99\%$ reflectance Spectralon reference, and then converted to absolute reflectance. The results for different PTFE suppliers are shown. The measurement values and uncertainties are described in Section~\ref{sec:spm_results_section}.}
\label{fig:spm_main_result}
\end{figure}

A summary of all the different reflectances measured with the UMS for the TPB-coated PTFE disks is presented in Figure~\ref{fig:spm_main_result}. The results are for 450~nm light in order to directly compare with the second method, where an LED of this wavelength is used (see Section~\ref{ssec:method3},~\ref{sec:comparison}). No results for UV wavelength are presented in this section, as the output signal of the UMS is interpreted as being purely due to reflectance, and thus is not accurate for the disks coated with TPB where there is significant fluorescence at that wavelength. For the uncoated PTFE disks, there is no clear trend for reflectance in thickness, so the results are consistent with previous work~\cite{Ghosh:2020}.

Since the measurements with the TPB-coated disks could only be performed once to avoid potential damage to the coating when inserting the disk in the integrating sphere, the data points shown in Figure~\ref{fig:spm_main_result} come from a single measurement. The error bars were estimated by repeating the measurement of a single uncoated disk with corresponding thickness twenty times. The magnitude of the error is not dependent on thickness, so this error is largely a systematic uncertainty from the UMS.

At 450~nm, for TPB-coated PTFE, the differences in reflectance between thicknesses measured are of the order of a $\sim$1--2\%, indicating that the reflectance of TPB-coated PTFE for blue light is not strongly dependent on the thickness of the material. For all the TPB-coated PTFE samples, the reflectance does not appear to monotonically increase with thickness, giving an indication that the primary factor for the observed differences is differences in sheet quality rather than variation with thickness. In addition, the reflectance of the ePlastics samples is consistently lower, indicating again that sheet quality would have a larger impact than thickness. Little change is observed between coated and uncoated Altec PTFE relative to our error bars, and this is as expected: the TPB should not significantly impact reflectance at 450~nm. These results agree with previous work, where the reflectance of uncoated PTFE was found not to depend on thickness in both~\cite{Haefner:2016} and ~\cite{Ghosh:2020}.

\section{Fluorescence spectrophotometer} \label{ssec:method2}

\begin{figure}[tb]
\centering
\includegraphics[width=0.49\textwidth]{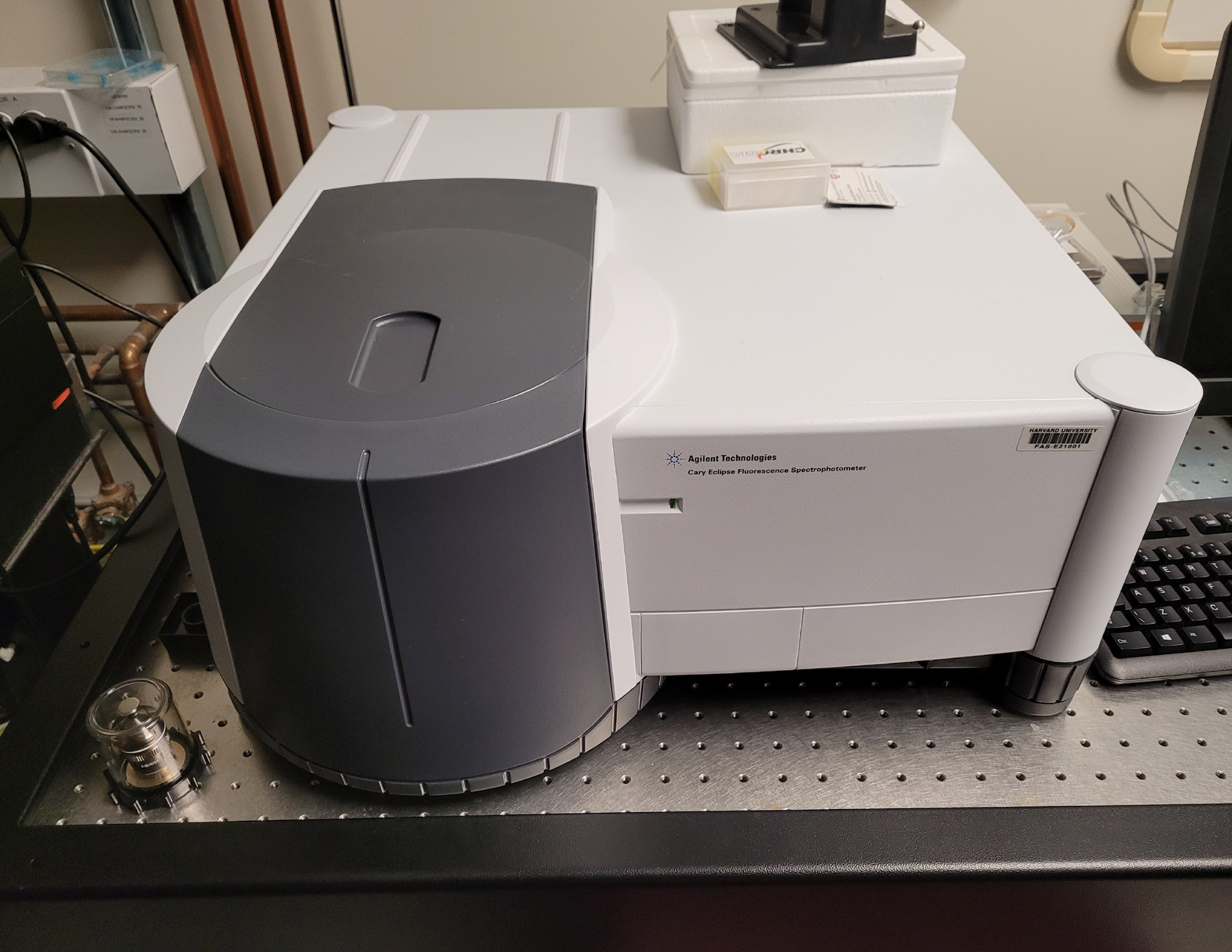}
\includegraphics[width=.49\textwidth]{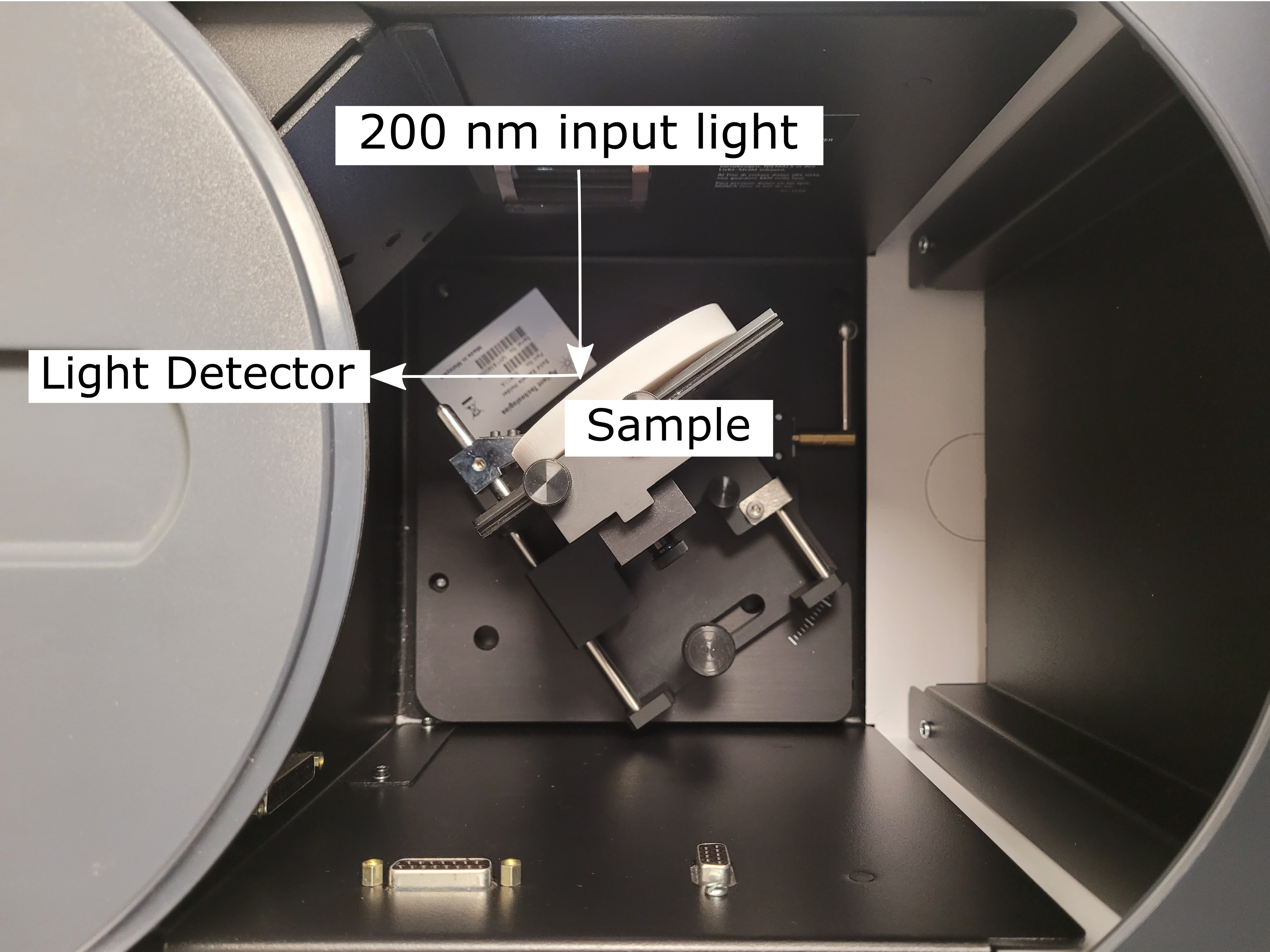}
\caption{Left: Picture of the \textsc{Cary Eclipse Fluorescence Spectrophotometer}. Right: The inside of the Fluorescence Spectrophotomer, with labels for the key components, where the sample is a disk of PTFE.}
\label{fig:fs_setup}
\end{figure}

The second method for determining light response uses a \textsc{Cary Eclipse Fluorescence Spectrometer} (FS),\footnote{https://www.agilent.com/cs/library/brochures/5990-7788EN\_Cary\_Eclipse\_Brochure.pdf} shown in Figure~\ref{fig:fs_setup}. The FS allows for the measurement of the fluorescence spectrum of a material for a range of wavelength inputs.

The fluorescence of the same TPB-coated disks described in section~\ref{ssec:method1} are measured using the FS. The error on these measurements are estimated by repeating the measurements in the FS 13 times for each of the four disks. Here, the FS is used to measure the wavelength at a fixed input wavelength of 200~nm. This is the lowest wavelength from the FS that still had a sufficient signal output for the purposes of comparison. The spectra for the four measured \textsc{Altec Plastics} disks for 200~nm light are shown in Figure~\ref{fig:fs_main_result}. Longer wavelengths (between 420~nm and 450~nm) were also studied in the FS, and it was confirmed that no fluorescence of TPB is observed for the blue light.

\subsection{Results from the fluorescence spectrophotometer}
\label{sec:fs_results_section}

\begin{figure}[!htb]
\centering
\includegraphics[width=0.8\textwidth]{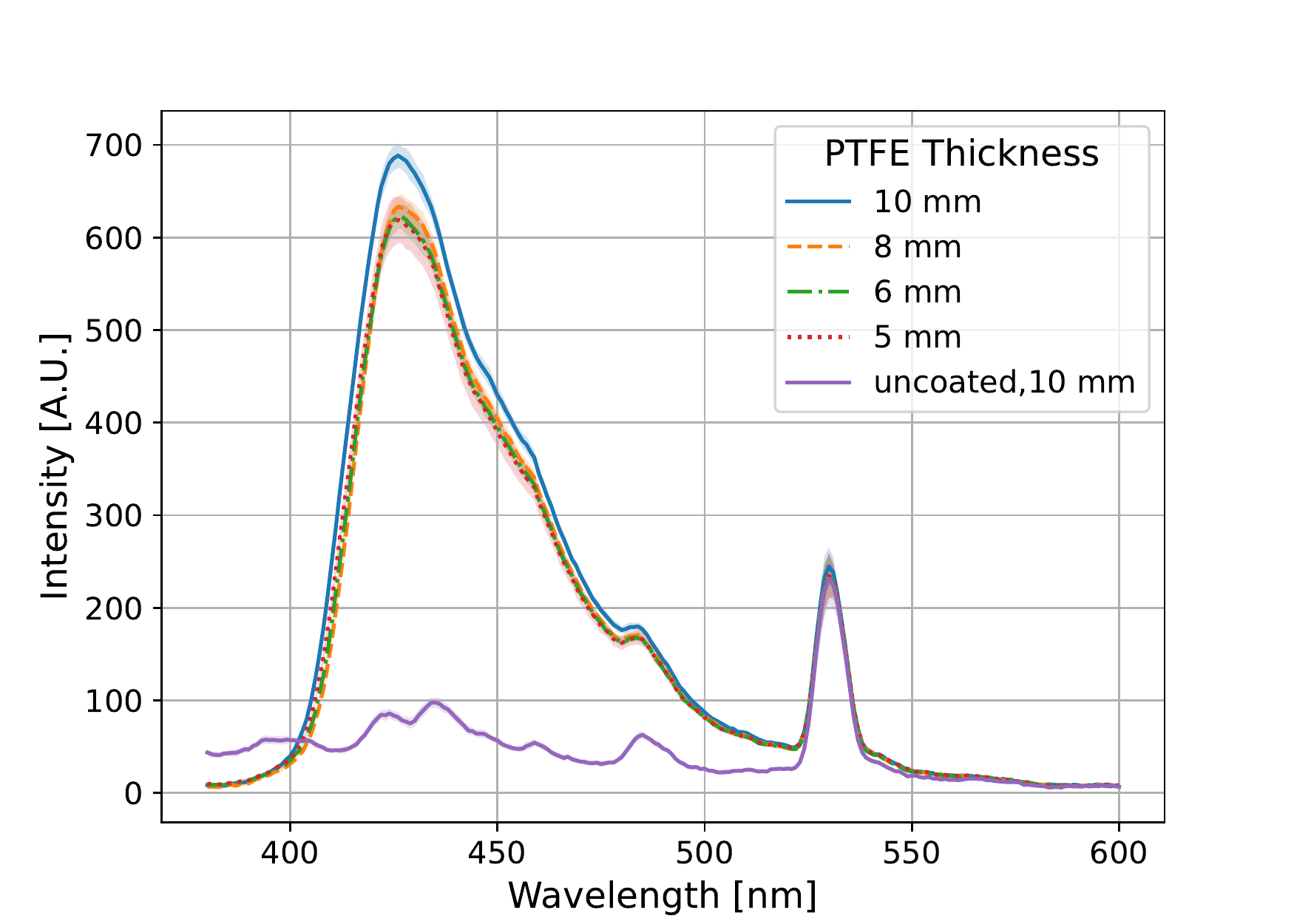}
\caption{Fluorescence spectrum measurements from the Fluorescence Spectrometer for different thicknesses of TPB-coated \textsc{Altec} PTFE (blue, orange, green, and red lines) and for an uncoated 10 mm thick disk (shown in purple). Uncertainties come from standard deviation of measurements taken 13 times for each disk.}
\label{fig:fs_main_result}
\end{figure}

When excited by UV light, TPB is expected to produce a fluorescence peak at around 430~nm \cite{Benson:2018}. The measured spectra from the TPB-coated \textsc{Altec} PTFE shown in Figure~\ref{fig:fs_main_result} clearly exhibit the expected peak. The measurements indicate that variations in fluorescence are not significant from 5-8~mm for the selected samples of \textsc{Altec} PTFE coated with TPB, but that the 10~mm thick disk fluoresces slightly more than the rest. The fluorescence of the 10~mm thickness is at most 15\%--20\% higher. This difference is likely not due to PTFE thickness, as the same result was not seen in the UMS measurements in Section ~\ref{ssec:method1}, but potentially to a variation in thickness of the TPB coating as a variation in WLSE of up to 14\% can be seen in Ref.~\cite{Benson:2018} when the variation in TPB thickness is up to 0.4~$\mu$m. This difference in TPB coating thickness was caused by a combination of factors: firstly, there is variation across each individual disk, which led to a measured difference of up to 10\% in intensity, in addition to coating variation across different evaporator runs, addressed in Section \ref{sec:Measurements}.

The peak at 530~nm has been observed as part of the TPB spectrum with light excited by 200~nm before and is a result of the interaction between diffraction effects in the FS and our sample, but not the TPB or PTFE \cite{Araujo:2019}. This hypothesis is supported by the fact that the peak is also visible for an uncoated PTFE disk, demonstrating that it is not caused by the TPB. Furthermore, a sample of plain white paper also shows that peak, discarding the origin being the PTFE itself. The only sample which did not produce such a peak was the Spectralon reference, which is engineered to have no fluorescence for any wavelength.

In conclusion, the FS measurements allow us to investigate exclusively the visible light that has been shifted by the TPB, rather than all the different wavelengths of light like in the UMS. These measurements have shown that the 5~mm, 6~mm and 8~mm disks of PTFE fluoresce near identical amounts of light, while the 10~mm disk produces a larger signal. The increase in signal from the 10~mm disk could be caused by variations in TPB thickness, rather than the thickness of the PTFE, either by evaporations, or by spot to spot variations on each disk. Either way, this underlines the importance of the box measurements in section ~\ref{ssec:method3}.


\section{PTFE boxes in air} \label{ssec:method3}

\begin{figure}[tbh!]
\centering
\includegraphics[width=0.7\textwidth]{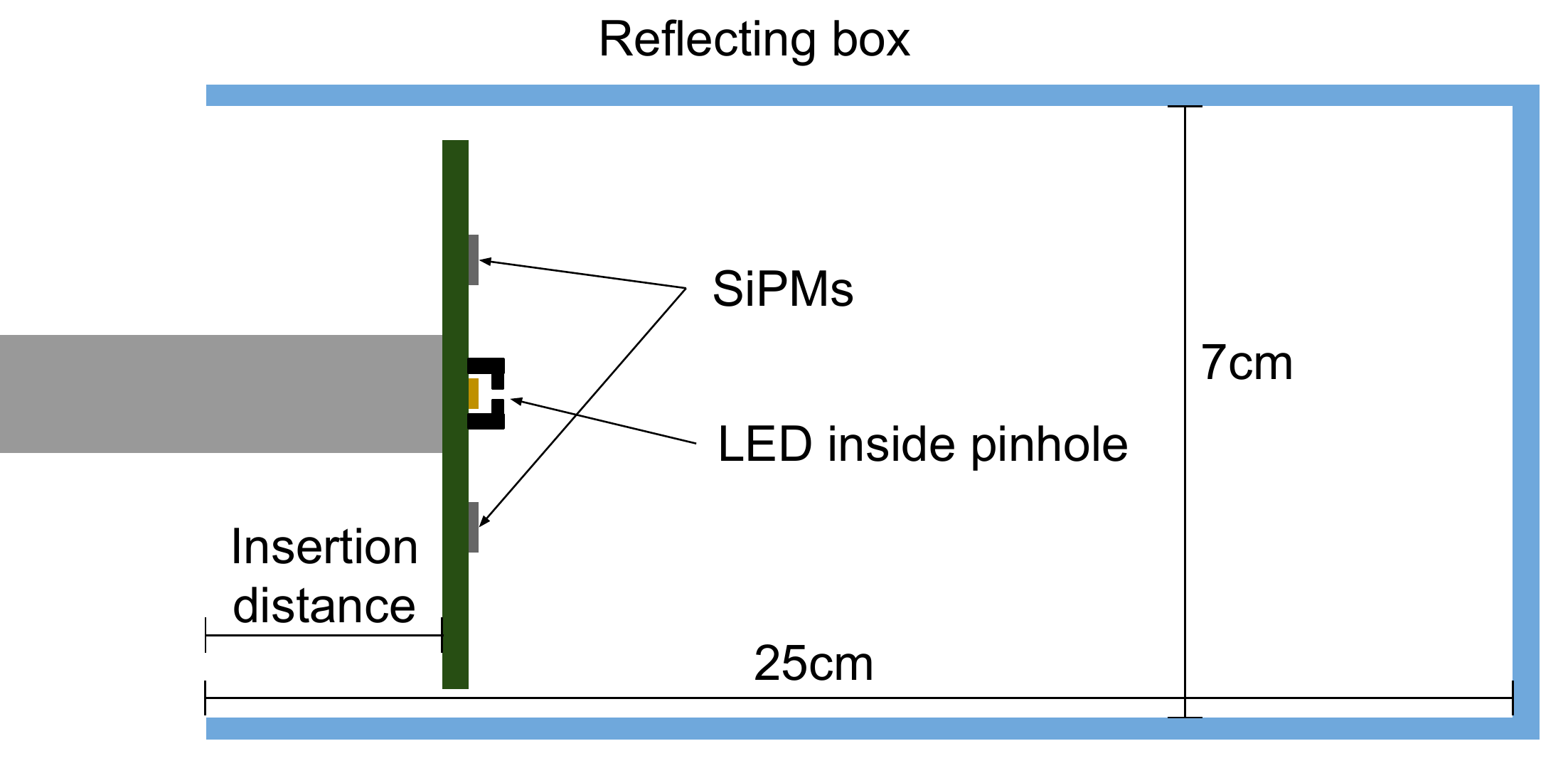}
\caption{Schematic side-view of the box setup. The inside dimensions of the box are $7~\mathrm{cm}~\times~7~\mathrm{cm}~\times~25~\mathrm{cm}$. The board holding the LED and SiPMs system is inserted through the open end of the box with a motorised rod. The amount of light collected is measured as a function of the distance of the board insertion to extract the reflectance of the PTFE box.}
\label{fig:box_setup}
\end{figure}

One significant disadvantage of the spectrophotometer methods (both UMS and FS) described in Sections~\ref{ssec:method1}~and~\ref{ssec:method2} is that the small area of the piece that is measured (approximately 0.2~cm$^2$) makes the measurements highly sensitive to potential variation across the piece or even across the sheet from which they were cut. Furthermore, the measurements do not allow extension into the VUV as the lowest wavelengths studied for the UMS is 260~nm and 200~nm for the FS. An alternative method described in this section, referred to as the \emph{box method}, uses a hollow rectangular prism of PTFE with a total interior surface area of 749~cm$^2$, illustrated in Figure~\ref{fig:box_setup}. This method allows to average potential reflectance variations within the larger PTFE samples by extracting the reflectance using detailed simulations of the setup. In Section~\ref{ssec:method4}, this method is also used in an argon environment to extend the studies to VUV wavelength.

\subsection{Box method setup}

The box method uses several rectangular boxes with inner dimensions of $7~\mathrm{cm}\times7~\mathrm{cm}\times25~\mathrm{cm}$ made of PTFE pieces of a given thickness, which were cut from the same \textsc{Altec} sheets used in the disk measurements. The boxes are open at one end to allow the insertion of a printed circuit board (PCB) with an LED and 4 silicon photomultipliers (SiPMs) that are sensitive to blue and UV light. The board is mounted on a motorized sliding rod such that the insertion distance can be changed quickly and precisely between measurements. A cross section of the setup is illustrated in Figure ~\ref{fig:box_setup}. Measurements of the light collected as a function of the distance of insertion are used to extract the reflectance for the uncoated case, whereas in the TPB-coated case, we measure the overall light signal (coming from reflection and fluorescence of the TPB). This process is further elaborated upon in Section \ref{sssec:Extraction of the reflectance value}. Other than minor improvements such as motorizing the system, this setup is largely identical to the experimental setup described in previous work ~\cite{Ghosh:2020}. In particular the dimensions of the boxes, the LEDs and the SiPMs are the same. However, in this work, the measurements are taken for insertion distances ranging from 0~cm to 16~cm at intervals of 2.3~cm compared to the previous work's 0~cm to 21~cm in 3~cm intervals. This change was made due to the physical improvements, such as motorizing the setup.

The systematic errors of these measurements were determined by measuring the variance in signal for different orientations of the box, to average out potential variations of the PTFE and/or TPB efficiencies for different parts. The box was made of 5 pieces, 4 identical rectangular pieces and 1 square shaped piece, with dimensions outlined above. When estimating the systematic errors of each box (uncoated and TPB-coated), the box is rotated in the 4 possible orientations relative to the PCB board. After this, the back piece was removed and placed on the other end, with the PCB now being inserted from the other side. Measurements in the 4 different orientations were also repeated for this reversed box configuration, summing to 8 total configurations being measured to estimated systematics variations at each insertion distance. Finally, in order to minimize temperature variations, measurements were taken within a single week, in a lab controlled to have temperature variations under $1^{\circ}$C.

\subsection{Extraction of the reflectance value}
\label{sssec:Extraction of the reflectance value}

In order to estimate the reflectances from the box measurements, the measurements are compared to detailed simulations of the setup with different reflectance values.
The simulations are performed using \textsc{Geant4} \cite{Allison:2016lfl}, where the box geometry includes the PTFE with a diffuse reflectance component only. The simulations are described in more details in previous work~\cite{Ghosh:2020}.

The simulations provide the total number of LED photons that reach the SiPMs based on the simulated PTFE reflectances. The simulation results for the different reflectances are fitted to the data where the reflectance value that best fits that data set is extracted. In order to allow comparison of the data taken with the SiPMs (in V$\cdot$ns) to the simulation (in photon counts), it is assumed that there is a constant and uniform conversion factor between the number of photons collected and the signal read out by the SiPMs, referred to as the scale factor parameter ($\alpha$), which is left free to vary in the fit. This $\alpha$ parameter incorporates effects such as the intensity of the LED and the efficiency of the SiPMs and should not change between measurements of different reflectances. Hence, the fit is performed simultaneously on all the data sets for the different reflectances using the same $\alpha$ parameter.

In the case of TPB-coated boxes in UV light, fluorescence becomes significant contributor to the overall light signal. To quantify this and to better compare different boxes, the previously described simulation is fit to the data in the same way, but the output metric becomes reflectance* instead of pure reflectance. Reflectance* is a combination of contributions from the reflected light and from the fluorescence light, and can be used to compare the overall light intensity, which retains a significant dependence on the PTFE reflectance. 

\subsection{Results of the box method}

\begin{figure}[tb]
\centering
        \includegraphics[width=.5\textwidth]{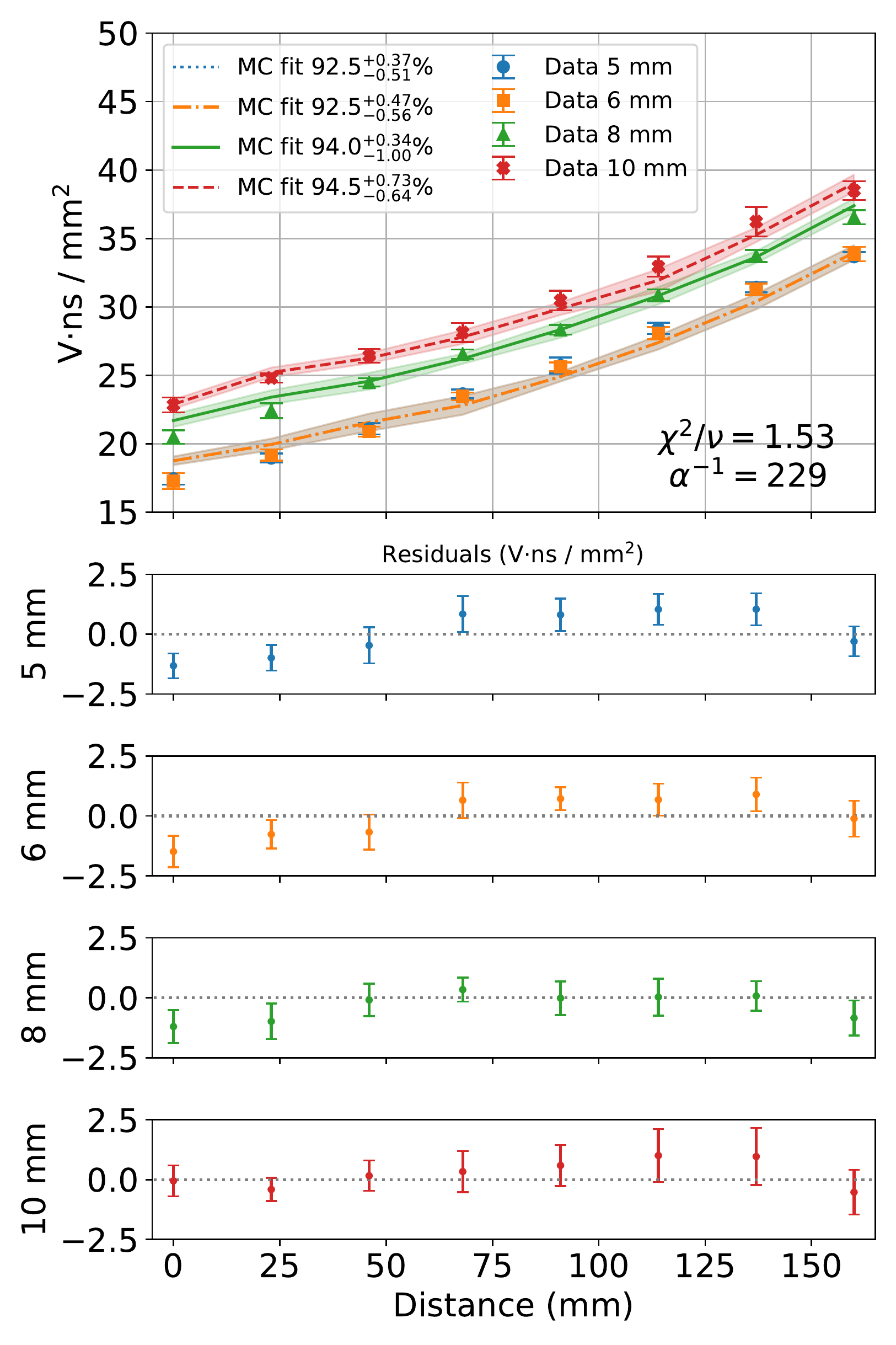}
        \includegraphics[width=.5\textwidth]{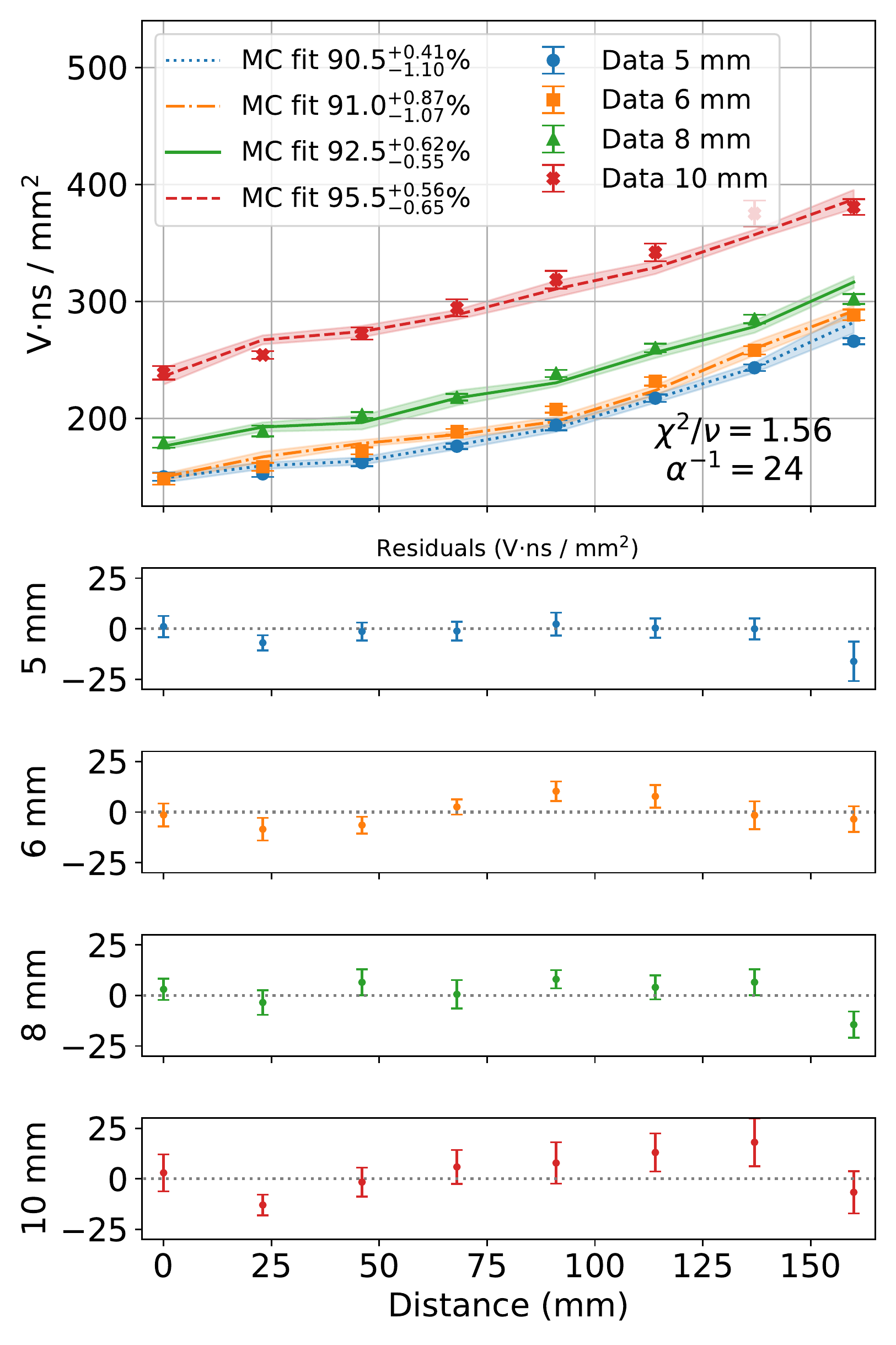}
\caption{Top plot: Measured signal (points) as a function of insertion distance for uncoated (left) and TPB-coated (right) PTFE boxes measured in air. On the left the setup used a blue (450~nm) LED, while on the right a UV (260~nm) LED is used. The lines represent the results of the simulations described in Section~\ref{sssec:Extraction of the reflectance value} which the points are then best fit to and the values in the legend provide the extracted reflectances(*) for each fit with the fit errors. The vertical axis gives the amount of light, measured in V$\cdot$ns by the SiPMs. The errors on the data points include statistical uncertainties of the SiPM pulses, as well as systematic errors regarding the rotations of the boxes. The errors on the simulations are statistical only~\cite{Ghosh:2020}. Bottom: These plots show the residuals between the data points and their corresponding simulated points (simulation minus data) for each of the insertion distances.}
\label{fig:combined_box_fit}
\end{figure}

Measurements of reflectances using the boxes were taken for the previously described thicknesses of 5~mm, 6~mm, 8~mm, 10~mm of \textsc{Altec} PTFE. Figure~\ref{fig:combined_box_fit} shows the results obtained with the 450~nm LED before the boxes were TPB-coated on the left, and shows the results obtained with the 260~nm LED for TPB-coated boxes on the right. This is done in an attempt to directly compare the impact of the PTFE thickness only, as once the 200~nm light has interacted with the TPB, the light reaching the PTFE is mostly at 430~nm and can be more directly compared to the blue LED shining on bare PTFE. For the 450~nm LED case, the fit to simulation provides a reflectance associated with each thickness. As already discussed, a direct reflectance value cannot be easily extracted for TPB-coated PTFE, as the fluorescence of the TPB adds to the reflectance and here reflectance* is computed.


\begin{figure}[tb]
    \centering
    \includegraphics[width = 0.8 \textwidth]{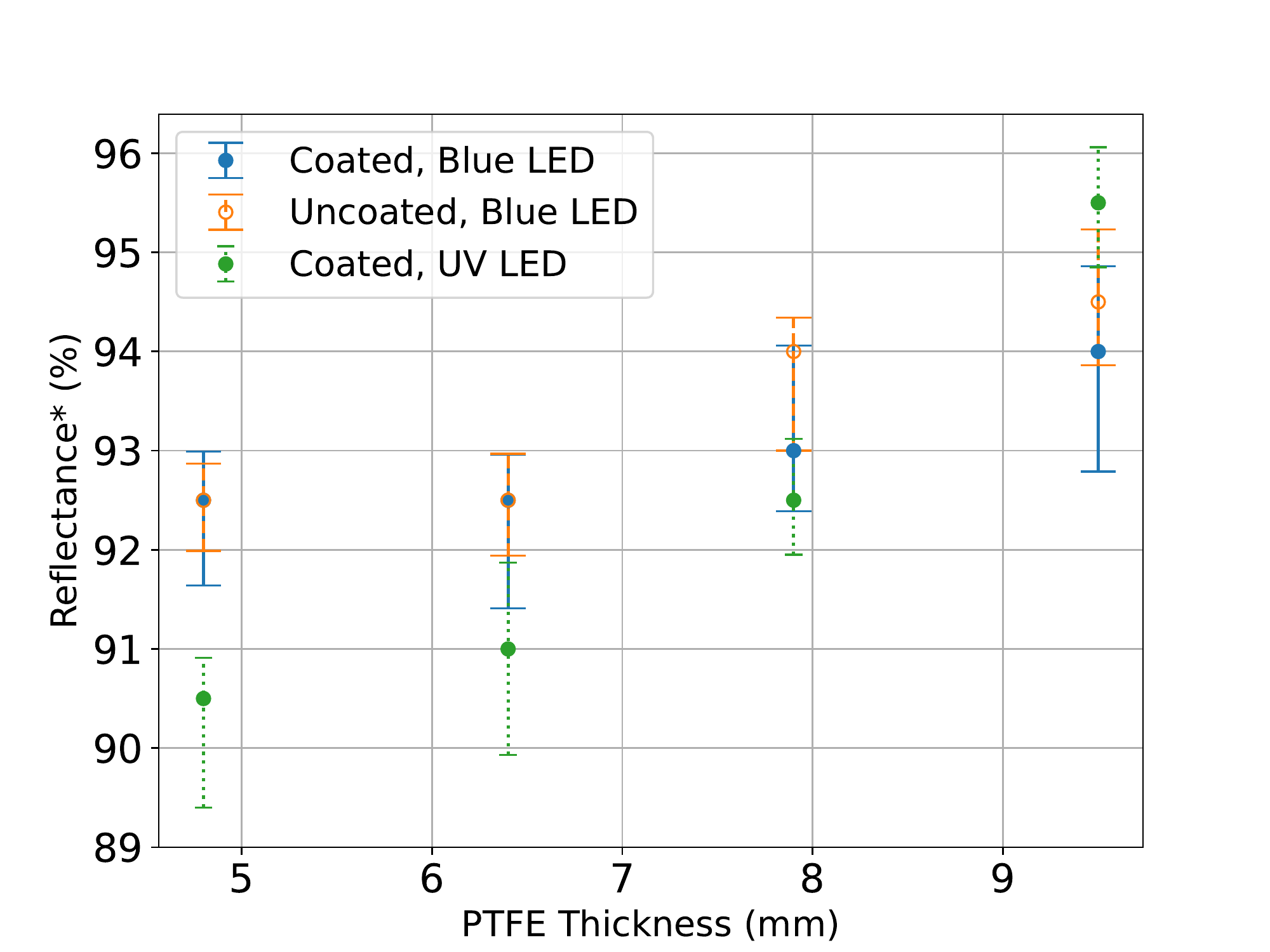}
    \caption{Comparison of reflectance values for uncoated and TPB-coated PTFE of different thicknesses measured with the box method for 450~nm and of reflectances* for 200~nm. The values for the coated measurements come from Figure~\ref{fig:combined_box_fit}. Note that the comparison between blue and UV light is only relative, as the values extracted for UV cannot directly translate to reflectance values.}
    \label{fig:CoatedUncoatedComp}
\end{figure}
 

In Figure~\ref{fig:combined_box_fit}, it can be seen that the simulations reproduce the data relatively well. The results for the 450~nm LED  are consistent with those previously reported in~\cite{Ghosh:2020}, though the previous measurements were with a different PTFE manufacturer (\textsc{ePlastics}). Differences in the MC fit reflectances of a few percent (2\% at most) are observed for the uncoated \textsc{Altec} PTFE pieces, and 5\% in the coated case.

To ensure that there is no additional effects produced by the TPB coating, a direct comparison of reflectance, using the blue (450~nm) LED, is performed on uncoated and TPB-coated PTFE boxes. Figure \ref{fig:CoatedUncoatedComp} shows that there is excellent agreement between the reflectances values extracted for uncoated and TPB-coated PTFE at 450~nm, indicating that the TPB does not, as expected, produce fluorescence at this wavelength. These directly agree with the conclusions of the UMS measurements in Section~\ref{ssec:method1}.

For 260~nm light, Figure~\ref{fig:combined_box_fit} shows some difference in signal strength with thickness. The 10~mm PTFE has the largest signal strength, followed by 8~mm, 6~mm and 5~mm, which are all close to one another (5~mm and 6~mm being approximately the same). This matches the results seen in the FS from Section~\ref{ssec:method2}, where the fluorescence of the 10~mm PTFE was larger than the rest, and 5~mm, 6~mm, and 8~mm were all close to one another.  These results are summarized in Figure~\ref{fig:CoatedUncoatedComp}, which shows the reflectance* values for different thicknesses of PTFE, compared to the results at 450~nm. Viewed in terms of this metric, while the curves are distinct (with the exception of 5~mm and 6~mm), the difference in reflectance* value does not reach more than 5\%. The reason the 10~mm box has a much larger reflectance* is likely due to the TPB coating and variations in runs of the evaporator, rather than the PTFE. This is similar reasoning as the measurement in the FS, found in Section~\ref{ssec:method2}.


In contrast to the UMS and FS measurements, the box measurements seem to indicate a stronger trend of reflectance with thickness. This demonstrates the value of having the two distinct methods. However, even for the box measurements the reflectance and reflectance* values have a difference of under 5\% for different thicknesses. In the measurements with the blue LED, the results are consistent with the results from the UMS, where coated and uncoated boxes have reflectance values that agree within error. In the measurements with the UV LED, the results agree with the results of the FS, where the 10~mm box stands out in comparison to the rest, but the magnitude of this difference remains small for NEXT. This could be caused by the fact that 10~mm pieces were coated together, and if that coating deposited more TPB, both the 10~mm box and disk would have higher reflectance. This relationship will be further investigated in a future work.  
\section{PTFE boxes in argon} \label{ssec:method4}

All measurements so far have been for light at a minimum of 200~nm, but in a typical gaseous xenon or argon experiment, the wavelength produced by scintillation would be even shorter (175~nm and 128~nm, respectively). Measuring the reflectance using the box method (described in Section~\ref{ssec:method3}) in gaseous argon with an alpha source, would thus allow us to verify the results of our light response measurements for TPB-coated PTFE with argon scintillation light.

\subsection{Argon chamber setup}

\begin{figure}[tbh]
\centering
\includegraphics[width=.4\textwidth]{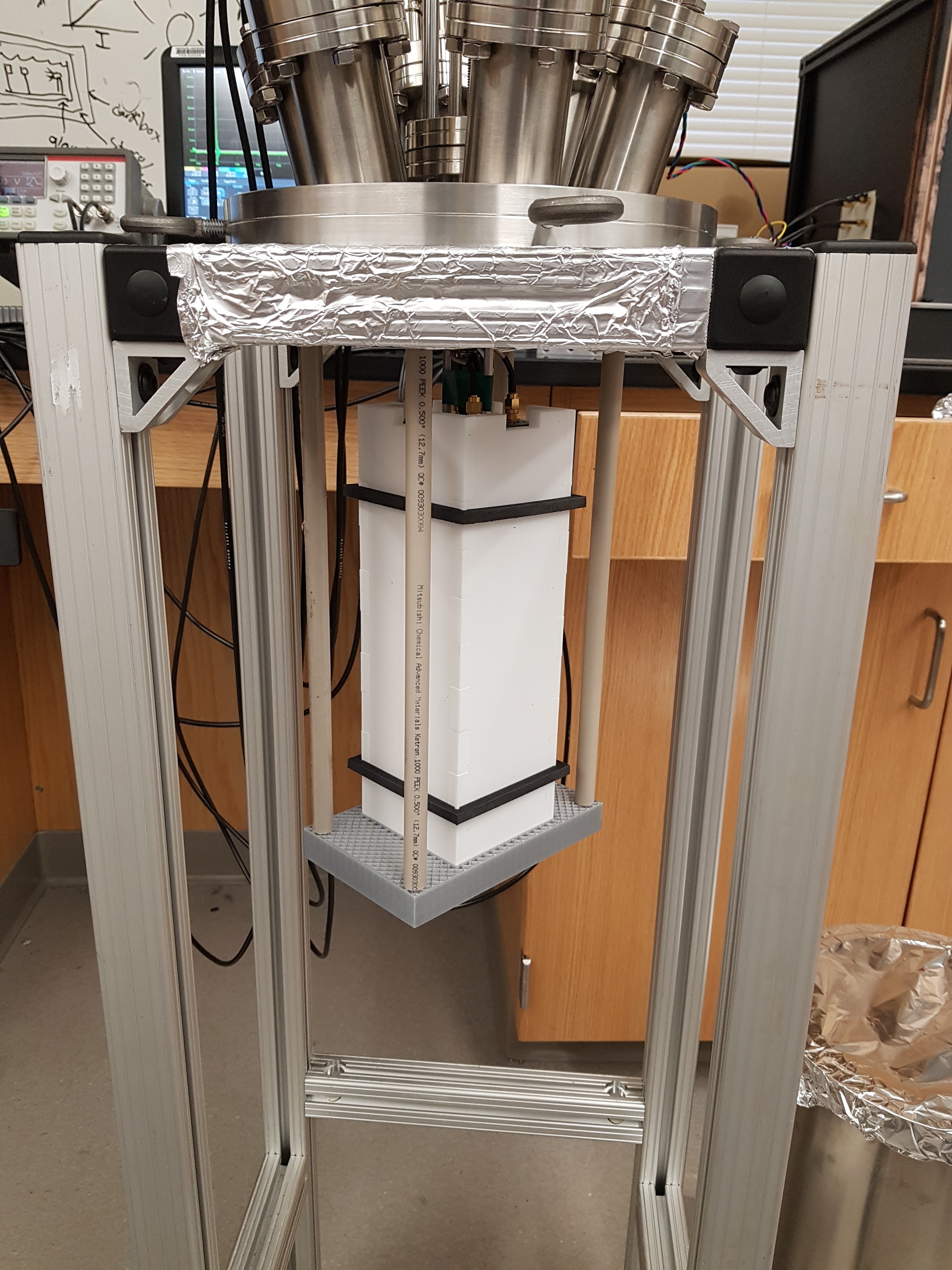}
\includegraphics[width=.38\textwidth]{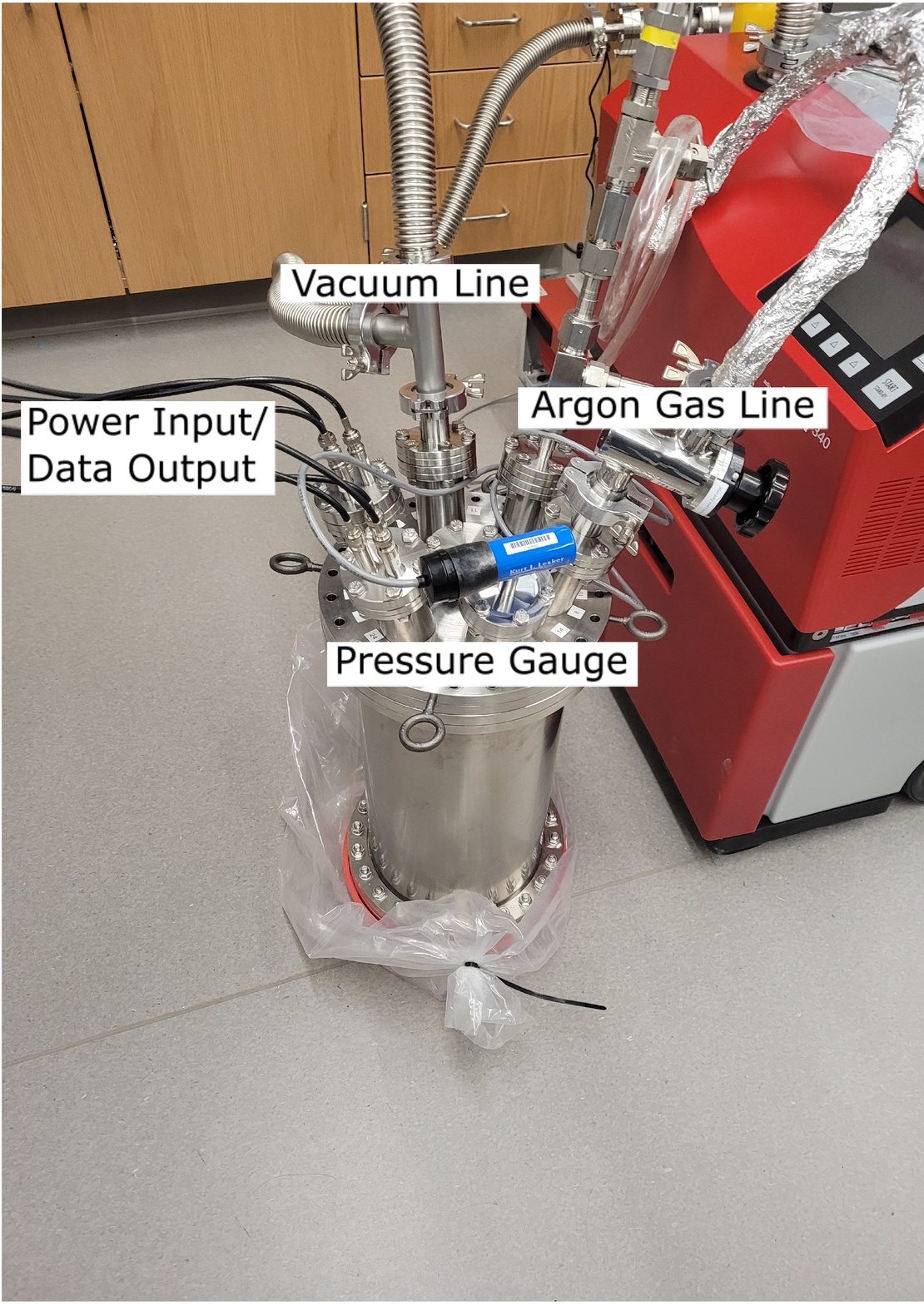}
\caption{Left: Picture of the argon chamber setup for a measurement. The main body of the chamber is removed so that the PTFE box and PCB holder are visible. Right: Picture of closed argon chamber, ready for measurements in a gaseous argon environment with radioactive source. }
\label{fig:ar_chamber}
\end{figure}

A setup identical to that described in Section~\ref{ssec:method3} is used for the measurement of TPB-coated PTFE boxes in argon. The LED source on the SiPM board is replaced with a $^{210}$Po source alpha emitter with an activity of 5~nCi~(185~events~$\cdot$~s$^{-1}$) supplied by \textsc{Spectrum Techniques}. 
The measurements were conducted in a sealed chamber filled with gaseous argon at approximately 1 bar of pressure at room temperature, shown in Figure~\ref{fig:ar_chamber}. It is expected that the alpha particle will travel about 6~cm from the source. Due to technical limitations of the setup, measurements of the light signals for the four boxes described in Section~\ref{ssec:method3} were taken at only two insertion distances, 6.0~cm and 13.7~cm. 

Approximately 6500 triggers are collected for each boxes at the two insertion distances, in 20 minutes runs. For each triggered pulse, full waveforms for all four SiPMs are saved. The small number of events relative to the source activity indicates that a large majority of alpha particles emitted by the $^{210}$Po source are not detected, mostly due to limitations from the trigger rate of the scope to read out and save full waveforms for all four SiPMs.

All events with nonzero signals in all four SiPM channels are considered to be alpha decays. For a given configuration (insertion distance + PTFE thickness), the measured signal is taken to be the sum of the median integrated signals (in all four channels) for all selected events in that configuration. The lower (upper) error bar corresponds to the 16$^{\text{th}}$ (84$^{\text{th}}$) percentile of the distribution of summed integrated signals, which result to be asymmetric. For a Gaussian distribution of event energies, this would correspond to the standard mean and standard deviation of the distribution.

\subsection{Results from PTFE boxes in argon}
\label{sec:argon_results_section}

Measurements of the signal as a function of distance for the \textsc{Altec Plastics} boxes were taken for the four thicknesses and are shown in Table~\ref{tab:ar_results}. 

The replacement of the LED with an alpha source makes it more challenging to assess the reflectance as a function of thickness with simulations in the same way as Section~\ref{ssec:method3}. Many additional factors need to be taken into account in the simulations, including but not limited to: the purity of the argon and the nitrogen contamination for light quenching affects, the argon pressure, and the quantum efficiency of TPB and SiPMs at the different wavelengths generated. Due to this difficulty, the method described above is not able to extract reflectance (or reflectance*) values, but is rather used to cross check that conclusions hold at 128~nm.  

Another limitation of the setup was the degree of nitrogen contamination. Due to some issues with the vacuum system, namely the vinyl tubing used to get the gaseous argon to the measurement setup, some level of air contamination occurred. Without a purity monitor it is unclear to what degree the contamination would change the signal but it is very likely a contributing factor to the error in both the size and number of signals. Because of this, these results should be taken qualitatively. In a future work, measurements into the degree of Oxygen and Nitrogen contamination would be required. 

\begin{table}[!htb]
    \centering
    \caption{A summary of median signal sizes (A.U.) of alpha decay events in TPB-coated PTFE boxes placed inside a gaseous argon enviroment (128~nm). Errors on each point range from the 16$^{\text{th}}$ to 84$^{\text{th}}$ percentile.}
    \begin{tabular}{ccccc}
        \toprule
        Insertion Distance &  \multicolumn{4}{c}{Sample thickness} \\
          & 5~mm&6~mm&8~mm&10~mm\\
        \hline \\
        6~cm & $34.1^{+5.8}_{-4.7}$ & $32.1^{+5.2}_{-4.2}$ & $30.1^{+4.8}_{-4.0}$ & 
        $29.7^{+4.5}_{-3.7}$ \\
        13.7~cm & $28.8^{+3.5}_{-3.7}$ & $34.2^{+5.4}_{-4.5}$ & $29.3^{+4.0}_{-3.5}$ & $31.4^{+4.5}_{-4.2}$ \\
    \end{tabular}
    
    \label{tab:ar_results}
\end{table}
\section{Comparison between methods}
\label{sec:comparison}

Four distinct measurement methods for determining the light response of TPB-coated PTFE as a function of thickness were presented above. Using the UMS provided measurements relative to a Spectralon reference, this allowed us to confirm that the TPB does not significantly impact the reflectance in the blue wavelengths. Measurements with the FS allowed us to directly characterize the impact of TPB on the response of PTFE to UV light, and to verify that there is only a weak dependence on thickness. Using the box methods allowed us to integrate together larger surface areas, and verify that our results still hold in a noble gas environment for argon scintillation light (128~nm).

\begin{figure}[!htb]
\centering
\includegraphics[width=0.7\textwidth]{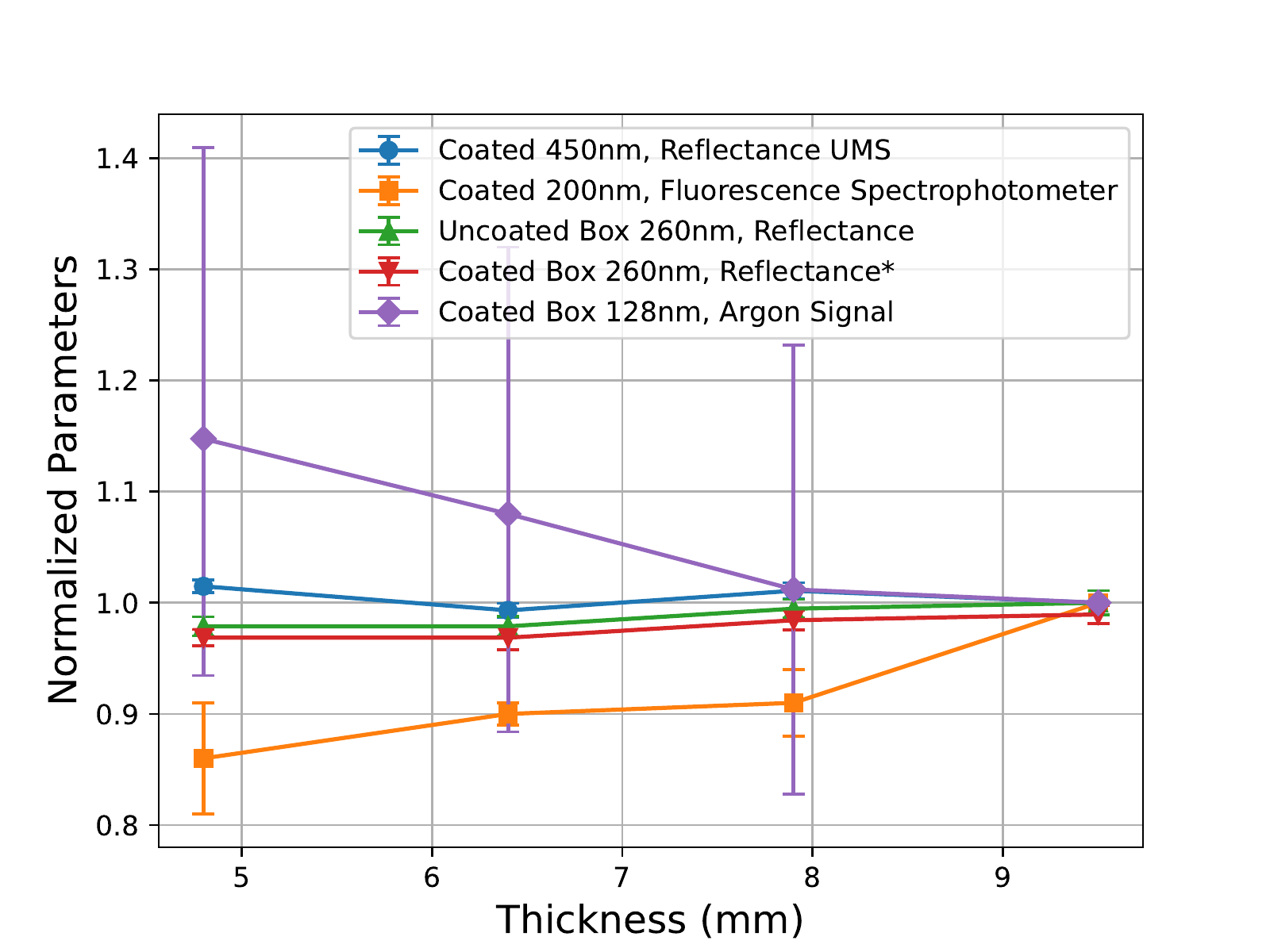}
\caption{Summary of all Altec PTFE results, combined in a single plot. For each distinct measurement method (UMS, FS, box in air and in argon) all data points are normalized by the corresponding value for the 10~mm box, allowing comparison between different methods. Some methods indicate minor differences between them, but overall results are consistent within error bars.}
\label{fig:results_summary}
\end{figure}

In order to compare between these distinct methodologies, the respective output values are normalized by the corresponding value for the thickest disk (10~mm). This makes it possible to display the variation in reflectance, fluorescence, and response to LED and argon scintillation light in one place, as shown in Figure~\ref{fig:results_summary}. In addition, a summary of the results are shown in Table \ref{tab:Summary}.
All of the results essentially point towards there being very little variation of TPB-coated PTFE light response with thickness in the range investigate (from 5~mm to 10~mm). The results all show reasonable agreement, although each draws out distinct features relating to the particularities of the measurement methods.
\begin{table}[!htb]
    \caption{A summary of the various results throughout the paper, all for plastics sourced from Altec Plastics. Descriptions of individual measurements can be found in the text. Measurement in argon taken at 6~cm insertion. Reflectance and reflectance* should not be directly compared.}
    \centering
    \def\arraystretch{1.0}%
    \begin{tabular}{cccccccc}
    \toprule
    \multirow{2}{*}[-2pt]{Method} & \multirow{2}{*}[-2pt]{\shortstack{Input\\ wavelength}} & \multirow{2}{*}[-2pt]{Coated} & \multirow{2}{*}[-2pt]{Units} & \multicolumn{4}{c}{Sample thickness} \\ 
        \cmidrule(lr){5-8}
        & & & & 10 ~mm & 8~mm & 6~mm & 5~mm \\
        \midrule
        UMS & 450~nm & y &  Refl. \% & $91.7 \pm 0.4$ & $92.7 \pm 0.5$ & $91.0  \pm 0.4$ & $93.0 \pm 0.3$\\
        FS & 200~nm & y & Norm. peak & 1.0 & 0.92 $\pm$ 0.02 & 0.90 $\pm$ 0.03 & 0.90 $\pm$ 0.04\\
        Box & 450~nm & n & Refl. \% & $94.5_{-0.51}^{+0.73}$ & $94.0_{-1.00}^{+0.34}$ & $92.5_{-0.56}^{+0.47}$ & $92.5_{-0.51}^{+0.37}$ \\
        Box & 450~nm & y & Refl. \% & $94.0_{-0.86}^{+0.49}$ & $93.0_{-1.09}^{+0.46}$ & $92.5_{-0.61}^{+1.06}$ & $92.5_{-1.21}^{+0.86}$ \\
        Box & 260~nm & y & Refl.* \% & $90.5_{-1.10}^{+0.41}$ & $91.0_{-1.07}^{+0.87}$ & $92.5_{-0.55}^{+0.62}$ & $95.5_{-0.65}^{+0.56}$ \\
        Ar & 128~nm & y & Arb. & $29.7^{+4.5}_{-3.7}$ & $30.1^{+4.8}_{-4.0}$ & $32.1^{+5.2}_{-4.2}$ & $34.1^{+5.8}_{-4.7}$\\
        \bottomrule
    \end{tabular}
    \label{tab:Summary}
\end{table}

\section{Conclusions and discussion}

Measurements of the variation in reflectance of TPB-coated PTFE with variations in thickness and wavelength (summarized in Figure~\ref{fig:results_summary}) were presented. It was found that the variation in reflectance at 450~nm for PTFE of thicknesses from 5~mm to 10~mm is measurable, ranging from 91\% to 93\% for our primary supplier, Altec Plastics. It was found that variation as a function of thickness at this wavelength for TPB-coated PTFE is very small, with an average variation of 2\% and no clear pattern with thickness, indicating a significant impact coming from the quality of the PTFE sheets themselves. Variations between suppliers were more significant, on average around 4\%. 

Similar consistency is found between the level of fluorescence light observed, with the samples under 10~mm all agreeing with each other to within error bars and the 10~mm approximately 14\% higher than the others. This increase is likely not caused due to the thickness of PTFE, but rather variation in the thickness of the TPB coating.

The measurements with PTFE boxes further reinforce these findings. The measurements in air clearly indicate that the different samples, while distinguishable, do not vary significantly. With the level of homogeneity observed in the NEXT White detector, NEXT-100 should be able to reach its target energy resolution~\cite{Lema:2018}. Furthermore, the measurements in gaseous argon give a good initial indication that our findings continue to hold for VUV light (128~nm), and in a detector environment at room temperature.

As the challenges faced by the NEXT experiment are common to many rare event searches (large mass of PTFE for light collection, need for low background), we anticipate these results will be valuable to the community. In particular, while PTFE can vary from vendor to vendor, our results indicate that reducing thickness can be beneficial for room temperature experiments. This is further confirmed by the consistency of the results for PTFE coated with TPB, and in an argon environment.

\acknowledgments

The NEXT Collaboration acknowledges support from the following agencies and institutions: the European Research Council (ERC) under Grant Agreement No. 951281-BOLD; the European Union’s Framework Programme for Research and Innovation Horizon 2020 (2014–2020) under Grant Agreement No. 957202-HIDDEN; the MCIN/AEI of Spain and ERDF A way of making Europe under grants RTI2018-095979 and PID2021-125475NB , the Severo Ochoa Program grant CEX2018-000867-S and the Ram’on y Cajal program grant RYC-2015-18820; the Generalitat Valenciana of Spain under grants PROMETEO/2021/087 and CIDEGENT/2019/049; the Department of Education of the Basque Government of Spain under the predoctoral training program of non-doctoral research personnel; the Portuguese FCT under project UID/FIS/04559/2020 to fund the activities of LIBPhys-UC; the Pazy Foundation (Israel) under grants 877040 and 877041; the US Department of Energy under contracts number DE-AC02-06CH11357 (Argonne National Laboratory), DE-AC02-07CH11359 (Fermi National Accelerator Laboratory), DE-FG02-13ER42020 (Texas A\&M), DE-SC0019054 (Texas Arlington) and DE-SC0019223 (Texas Arlington); the US National Science Foundation under award number NSF CHE 2004111; the Robert A Welch Foundation under award number Y-2031-20200401. Finally, we are grateful to the Laboratorio Subterr’aneo de Canfranc for hosting and supporting the NEXT experiment.

 

\providecommand{\href}[2]{#2}\begingroup\raggedright\endgroup

\bibliographystyle{JHEP}

\end{document}